\documentclass{article}
\usepackage{0_arxiv}

\usepackage[utf8]{inputenc} 
\usepackage[T1]{fontenc}    
\usepackage{hyperref}       
\usepackage{url}            
\usepackage{booktabs}       
\usepackage{amsfonts}       
\usepackage{nicefrac}       
\usepackage{microtype}      
\usepackage{fancyhdr}       
\usepackage{graphicx}       
\usepackage{lipsum}
\usepackage[ddmmyyyy]{datetime}
\newdateformat{mydate}{\twodigit{\THEDAY}{ }\shortmonthname[\THEMONTH], \THEYEAR}

\usepackage{amsmath} 
\usepackage{amssymb} 
\usepackage{dsfont}
\usepackage[pdftex,dvipsnames]{xcolor}
\usepackage[colorinlistoftodos,prependcaption]{todonotes}

\usepackage[babel,german=guillemets]{csquotes}
\usepackage{float}
\usepackage{graphicx} 
\usepackage{listings}
\usepackage{algorithm}
\usepackage[noend]{algpseudocode}
\usepackage{makecell}
\usepackage{microtype} 
\usepackage{mparhack}
\usepackage{multicol} 
\usepackage{multirow} 
\usepackage{mathtools}
\usepackage{paralist} 
\usepackage{wrapfig} 
\usepackage{physics} 
\usepackage[verbose]{placeins}
\usepackage{siunitx}
\usepackage[format=hang, font={small,up}, labelfont=bf, justification=justified, singlelinecheck=false]{caption}
\usepackage{subcaption}
\usepackage[nottoc]{tocbibind}
\usepackage{svg} 
\usepackage{my_newcommands}
\usepackage{algpseudocode}
\usepackage{algorithm}
\usepackage{cooltooltips}
\usepackage{wrapfig}

\title{Merging automatic differentiation and the adjoint method for photonic inverse design}

\author{
    Alexander Luce \\
    University Erlangen-Nürnberg \&\\
    ams OSRAM Group\\
    Regensburg \\
\And
    Rasoul Alaee \\
    ams OSRAM Group\\
    Regensburg\\
\And
    Fabian Knorr \\
    ams OSRAM Group\\
    Regensburg\\
\And
   Florian Marquardt \\
   University Erlangen-Nürnberg \&\\
   Max Planck Institute \\
   for the Science of Light\\
   Erlangen \\
}

\begin{document}
\maketitle
\begin{abstract}

Optimizing shapes and topology of physical devices is crucial for both scientific and technological advancements, given its wide-ranging implications across numerous industries and research areas. Innovations in shape and topology optimization have been seen across a wide range of fields, notably structural mechanics, fluid mechanics, and photonics. Gradient-based inverse design techniques have been particularly successful for photonic and optical problems, resulting in integrated, miniaturized hardware that has set new standards in device performance. To calculate the gradients, there are typically two approaches: implementing specialized solvers using automatic differentiation or deriving analytical solutions for gradient calculation and adjoint sources by hand. In this work, we propose a middle ground and present a hybrid approach that leverages and enables the benefits of automatic differentiation and machine learning frameworks for handling gradient derivation while using existing, proven solvers for numerical solutions. Utilizing the adjoint method, we turn existing numerical solvers differentiable and seamlessly integrate them into an automatic differentiation framework. Further, this enables users to integrate the optimization environment with machine learning applications which could lead to better photonic design workflows. We illustrate the approach through two distinct examples: optimizing the Purcell factor of a magnetic dipole in the vicinity of an optical nanocavity and enhancing the light extraction efficiency of a \textmu LED.
\end{abstract}


\section{Introduction}
\label{sec:1_introduction}

Optimization is a crucial aspect in the development of structural devices that dictate the physical properties of waves and fields in order to yield higher performance compared to those created using traditional approaches. The application domain for optimization for light is vast and rapidly evolving \cite{2023:bliokh:roadmap_on_structured_waves}, encompassing numerous techniques that modify parameters or geometries based on specific update algorithms. Generally, physical systems subject to optimization do not exhibit a convex loss landscape, resulting in inherent limitations when seeking global optima. Consequently, most optimization algorithms can be categorized as either global or local optimization.
Global optimization techniques for photonic optimization include evolutionary algorithms \cite{2018:Johlin:Broadband_highly_directive_3D_nanophotonic_lense} and Bayesian optimization techniques \cite{2022:Wankerl:Directional_emission_of_white_light_via_selective_amplification_of_photon_recycling_and_Bayesian_optimization_of_multi-layer_thin_films, 2021:Garcia:Numerical_methods_for_shape_optimization_of_photonic_nanostructures}. Although global optima are typically preferred over local optima, these algorithms face significant limitations regarding their applicability. For instance, Bayesian optimization becomes increasingly expensive for large problems with many parameters and data points \cite{2016:Springenberg:Bayesian_Optimization_with_Robust_Bayesian_Neural_Networks}. Similarly, evolutionary algorithms are often sample-inefficient and unsuitable for computationally expensive evaluations. Emerging machine-learning approaches offer new possibilities for global optimization and have demonstrated promising results in photonic optimization applications \cite{2020:An_Qing:multilayer_deep_q, 2020:Wankerl:Parameterized_Reinforcement_Learning_for_Optical_System_Optimization, 2020:So:Deep_learning_enabled_inverse_design_in_nanophotonics, 2017:peurifoy:Nanophotonic_Particle_Simulation_and_Inverse_Design_Using_Artificial_Neural_Networks, 2022:Yeung:Enhancing_Adjoint_Optimization-Based_Photonic_Inverse_Design_with_Explainable_Machine_Learning}. However, they cannot overcome the fundamental issue of the curse of dimensionality \cite{2017:Keogh:Curse_of_Dimensionality}. 
Conversely, identifying local optima in high-dimensional problems is much more manageable than obtaining a global minimum \cite{2018:Nesterov:Lectures_on_Convex_Optimization}, as evidenced by the vast number of neural network parameters being optimized during training for deep learning and machine learning  \cite{2023:Birhane:Science_in_the_age_of_large_language_models}. By utilizing gradient-based optimization and adaptive stepsize methods such as ADAM \cite{2017:kingma:Adam:_A_Method_for_Stochastic_Optimization} or line-search \cite{1987:Fletcher:Practical_Methods_of_Optimization}, it is feasible to optimize numerous parameters simultaneously and achieve a local optimum. Gradient-based optimization is often applied in numerical optimization algorithms for device parameter, shape, or topology optimization, which is commonly referred to as inverse design \cite{2018:Molesky:Inverse_design_in_nanophotonics, 2013:Miller:Adjoint_shape_optimization_applied_to_electromagnetic_design, 2015:Hansen:Accurate_adjoint_design_sensitivities_for_nano_metal_optics, 2020:Augenstein:Inverse_Design_of_Nanophotonic_Devices_with_Structural_Integrity, 2011:Jensen:Topology_optimization_for_nano-photonics}. To employ gradient-based optimization, it is essential to compute the gradients of the desired function with respect to a loss or target value. This computation can be challenging, as the loss often depends on the solution of the system of equations governing the underlying physical problem. The adjoint method \cite{2021:Johnson:Notes_on_Adjoint_Methods, 1987:Pontriagin:The_Mathematica_Theory_of_Optimal_Processes} allows for obtaining analytical gradients by deriving adjoint equations, which can then be manually integrated into numerical solvers and update equations for the geometry parameters \cite{2019:Lebbe:Contribution_in_topological_optimization_and_application_to_nanophotonics, 2015:Hansen:Accurate_adjoint_design_sensitivities_for_nano_metal_optics, 2013:Miller:Adjoint_shape_optimization_applied_to_electromagnetic_design, 2009:Durand:Adjoint_variable_method_for_time-harmonic_Maxwell_equations, 2013:Lalau:Adjoint_shape_optimization_applied_to_electromagnetic_design}.
However, this approach can be tedious and problem-specific, requiring new derivations for different physical settings or optimization targets.
Automatic differentiation (AD) offers an alternative by automating the complex and elaborate process of deriving gradients \cite{2017:Baydin:Automatic_Differentiation_in_Machine_Learning:_A_Survey}. With AD, one only needs to implement the forward function, provide a suitable parameterization, define functions for solving the governing partial differential equation (PDE), postprocess the PDE solution, and define the target/loss function. Although implementing functions for the postprocessing and the loss is typically straightforward, obtaining gradients of a physical solution for a PDE can be challenging with AD, especially when established numerical solvers do not support it. Consequently, an end-to-end approach using AD requires implementing a new solver directly within the AD framework to derive the PDE solution and perform backpropagation \cite{2019:Vuckovic:Nanophotonic_Inverse_Design_with_SPINS:_Software_Architecture_and_Practical_Considerations, 2023:Vial:Nannos, 2020:Minkov:Inverse_design_of_photonic_crystals_through_automatic_differentiation, 2019:Hughes:Forward-Mode_Differentiation_of_Maxwell’s_Equations}. This task can be impractical since migrating an existing, validated physical model to an AD solver and framework poses significant work overhead.

In this work, we propose a "hybrid approach" that merges the benefits of automatic differentiation (AD) with the applicability of the adjoint equation to established solvers. By utilizing analytically derived results from previous works on the physical adjoint problem, we directly integrate established numerical solvers into an AD framework. Importantly, the internal workings of the numerical solver must not be accessible to the user. Hence, we consider the adjoint computation to be an atomic step in the computational graph of the AD framework. This combination creates an end-to-end AD-enabled process incorporating established numerical solvers. The solvers are then seamlessly integrated into the computational graph of the AD framework, effectively rendering them auto-differentiable for the optimization. This approach allows users to leverage the functionality and efficiency of modern AD frameworks while selecting the optimal numerical solver for their specific problem, regardless of AD compatibility\footnote{It is crucial for the solver to provide an interface that enables loading external geometries or parameters, and sources, as well as exporting numerical solutions, which typically is a supported functionality \cite{lumerical, comsol}.}.

We demonstrate the application of AD integration by performing shape optimization on two distinct photonic problems of scientific and engineering interest. In the first example, we aim to enhance the Purcell factor of a photonic nanocavity by deforming the cavity geometry. In the second example, we apply shape optimization to the outcoupling structure of a µLED. In both cases, we implement the analytically derived equations for the shape gradient into PyTorch, while utilizing PyTorch-provided functionality for postprocessing and loss calculations \cite{2017:Paszke:Automatic_differentiation_in_PyTorch}.

\section{Combining the adjoint method with autodifferentiation}
\label{sec:adjoint_method_AD}

A general optimization problem for photonic applications is the overarching goal to achieve the highest possible value of a physical property. The problem is typically described by a partial differential equation (PDE) $A$ that governs the dynamics. Here, we consider only linear PDEs for brevity, but in general, the dynamics could also encompass non-linear equations or be described by an eigenvalue problem \cite{2021:Johnson:Notes_on_Adjoint_Methods}. With a set of geometrical design parameters $p$, the solution $u$ of the physical system is given by the equation $A_p u = b_p$ where $b_p$ denotes source terms. The system under consideration is embedded in a simulation region $\mathcal{D}$ on which the solution is computed. A figure of merit or loss $J(u)$ must also be defined which evaluates the solution of the physical system given by the parameters $p$. Typically, this loss functional is given by an integral over the computational domain $J = \int_\mathcal{D} u(s) \d s$ or a sum over discrete physical properties of the solution. The optimization problem is formulated by $\min_p J(u_p)$ such that $A_p u = b_p$. Typical examples are increasing the quality factor of a cavity, focusing the emission of light into a particular solid angle, or increasing the Purcell effect for an emitter \cite{2003:Vahala:Optical_Microcavities, 2018:Johlin:Broadband_highly_directive_3D_nanophotonic_lense}. The problem of computing the optical characteristics for a given task is usually tackled by employing various numerical solvers such as rigorous coupled wave analysis (RCWA), finite difference time domain (FDTD), finite-difference frequency-domain (FDFD), or finite element method (FEM). Depending on the type of problem, the approximations and discretizations performed by the software are particularly useful for a specific problem. As a result, many specialized types of solvers and software solutions exist today. However, the situation is usually not as straightforward as choosing a set of parameters that can be applied to the PDE and then be evaluated directly. For more complex problems, geometry generation and postprocessing of the raw solution of the PDE require additional work on top of finding the solution. The evaluation of the design of an optimization process can be separated into the following steps: the geometry definition \autoref{fig:sim_flow}a), the numerical simulation of the physical problem (\autoref{fig:sim_flow}b)), the postprocessing of the results (\autoref{fig:sim_flow}c)), and the evaluation by a loss functional (\autoref{fig:sim_flow}d)).

Optimizing a large set of parameters $p$ by gradient-based optimization requires the gradient of the loss functional with respect to the design parameters $\delta J(u)$. Obtaining gradients of a loss functional $J$ with respect to the input parameters $p$ is difficult at first glance since it involves computing the variation of the loss functional with respect to all input parameters. This would lead to a computational complexity that scales linearly with the number of input parameters. For large problems, this poses a heavy computational burden. Fortunately, the computational complexity can be improved to a constant dependency on the input parameters by either employing the adjoint method or using (backward mode) automatic differentiation to compute the parameter gradients. Both approaches are elucidated and combined in the following.

\subsection{Automatic differentiation}
Backward mode automatic differentiation or backpropagation is the idea of applying the chain rule algorithmically to a numerically executed computation. The gradient is separated into atomic computations for which the derivatives are known analytically. For a simple computation $y = f(g(x))$, the chain rule gives the derivative of y with respect to x as $\frac{\p y}{\p x} = \frac{\p f}{\p g}\frac{\p g}{\p x}$. The functions $f$ and $g$ can now be differentiated individually and the analytical solution of the respective derivative reused whenever the functions appear again. For arbitrarily big computations, it suffices to separate the computation into individual function applications and keeping track of the order in which all functions have been applied. This concept is known as the computational graph. Consider a complicated computational graph $J = f_G(p)$ that represents solving the linear PDE of the aforementioned physical system $A_p u = b_p$ and computing the loss of the system $J(u)$. Computing $\frac{\p J(u)}{\p p}$ is then reduced to tracing back the individual steps of the computational graph from the loss $J$ to the parameters $p$ and chaining the respective derivatives of the intermediate function applications. To distinguish which part of the computation is executed, any mathematical function of the AD framework provides a forward and a backward method. The normal function evaluation is applied when the forward method is invoked since the computational graph is traversed in the forward direction from the input to the loss. The derivative is computed by the backward method which receives the gradient information in backward order starting from the loss and propagating to the input values. Automatic differentiation is straightforward to use and many scientific AD frameworks are readily available such as Jax, PyTorch, Hips/autograd, and more \cite{2018:Bradbury:Jax_composable_transformations_of_Python_Numpy_programs, 2017:Paszke:Automatic_differentiation_in_PyTorch, 2015:maclaurin:autograd_Effortless_gradients_in_numpy, 2023:white:juliadiff, 2023:Leal:autodiff}. The drawback of using AD frameworks for optimization is that the numerical solver for the PDE must support automatic differentiation and create a computational graph during the numerical solution for $u$. Although this is a rapidly evolving field and a number of solvers employing automatic differentiation have been developed \cite{2023:Vial:Nannos, 2019:Vuckovic:Nanophotonic_Inverse_Design_with_SPINS:_Software_Architecture_and_Practical_Considerations, 2019:Hughes:Forward-Mode_Differentiation_of_Maxwell’s_Equations}, many popular choices and industry-standard solvers do not support AD \cite{lumerical, comsol}.

\subsection{Adjoint Method}
On the other hand, adjoint methods approach the gradient derivation from the manual side. Here, we briefly recall the basis of the adjoint method \cite{2021:Johnson:Notes_on_Adjoint_Methods, 2019:Lebbe:Contribution_in_topological_optimization_and_application_to_nanophotonics, 2015:Hansen:Accurate_adjoint_design_sensitivities_for_nano_metal_optics}. Consider again a system governed by the linear PDE $A_p u = b_p$ with loss functional $J$. The functional derivative of $J(u)$ can be expressed by $\frac{\d J}{\d p} = \frac{\p J}{\p p} + \frac{\p J}{\p u}\frac{\p u}{\p p}$.
The term $\frac{\p u}{\p p}$ exhibits the undesired effect of linear scaling of the computational cost with the number of input parameters $p$ if it is evaluated via finite differences. However, this term can be expressed by taking the derivative of the PDE wrt. to p 
$\frac{\p Au}{\p p} - \frac{\p b}{\p p} = \frac{\p A}{\p p}u + A\frac{\p u}{\p p} - \frac{\p b}{\p p} = 0$ which is than rearranged to 
\begin{align}\label{eq:pde_derivative}
    \frac{\p u}{\p p} = A^{-1}\left(\frac{\p b}{\p p} - \frac{\p A}{\p p} u\right).
\end{align}
Putting everything together, we are given an equation that splits into three parts - the derivative of the loss with respect to the parameter, the s.c. \textit{adjoint solution} and the \textit{forward solution}: 
\begin{align}\label{eq:adjoint_method}
    \frac{\d J(u)}{\d p} = \frac{\p J}{\p p} +  \underbrace{\left(\frac{\p J}{\p u}A^{-1}\right)}_\text{adjoint solution} \underbrace{\left( \frac{\p b}{\p p} - \frac{\p A}{\p p}u \right)}_\text{forward solution}. 
\end{align}
Here, the name adjoint solution derives from the reformulation of the left side of \autoref{eq:adjoint_method} to $A^\dagger v = \frac{\p J}{\p u}$ where $v$ is the solution of this adjoint PDE. For many linear PDE, the adjoint system equations $A^\dagger$ are straightforward to derive. The boundary terms are given by the sensitivity of the loss functional with respect to the solution of the original PDE. The right side of \autoref{eq:adjoint_method} is then comprised of the solution $u$ and the sensitivity of the system matrix $A$ and boundary conditions of the original PDE. The adjoint method is appealing due to its generality. It is applicable to physical optimization problems in many settings, also outside of photonics. In particular, it is possible to compute the forward and adjoint solution with many different types of numerical solvers as long as the adjoint system equations can be used by the solver. 
However, it poses an analytical overhead before the optimization since the derivation of the required equations is done by hand. Especially for complicated problems where the system equations have a complicated dependency on the parameters in $\frac{\p A}{\p p}$ and $\frac{\p b}{\p p}$ or if the loss functional involves lengthy and tedious postprocessing $J = f \circ g\circ h \dots$ the adjoint method becomes impractical. 

\subsection{Integrating the adjoint method into automatic differentiation}
Interestingly, the advantages and disadvantages of AD and the adjoint method seem to complement each other. While the adjoint method is difficult to use but generally compatible with most numerical solvers, AD is easy to use but only applicable by using appropriate solvers. Here, we show how to combine both methods and leverage the advantages of each other to cancel their disadvantages. 

The key idea is to incorporate the adjoint method directly into the computational graph of the AD framework. This integration provides the benefit of the AD framework's flexibility without the necessity of rewriting efficient numerical solvers. In order to integrate the adjoint method into the backward calculation of an AD framework, we need to identify the appropriate terms in the derivation. Fortunately, the forward and adjoint solution derivation is similar to the distinction between the forward and backward methods for AD. The forward method should receive the input parameters $p$ that determine the system equations $A_p$ and source terms $b_p$ and return the solution of the PDE $u$ back to the computational graph. During gradient computation, we crucially depend on the input from the backward methods to receive the \textit{adjoint source} $\frac{\p J}{\p u}$ with which the adjoint solution $v$ can be computed. In the backward method, we should therefore receive gradient information from the loss function and any other postprocessing steps that were computed from the forward solution $u$. Then the adjoint solution and the forward solution are multiplied (\autoref{eq:adjoint_method}) and the solution is returned back to the computational graph for further processing back to the root of the parameters. The remaining terms $\frac{\p A}{\p p}$ and $\frac{\p b}{\p p}$ depend on the discretization of the solver and the applied optimization scheme and must be treated accordingly if they are not part of the AD framework. 

There are two popular schemes that are mostly used for the geometric optimization of photonics - topology and shape optimization. In topology optimization, the entire distribution of material is considered to change point-wise throughout the optimization domain. In shape optimization, the boundary $\p\Omega$ of a shape $\Omega$ is continuously deformed but the shape remains connected during the deformation. Importantly, the approaches require different treatments of the gradients. We will focus on shape optimization in the following but similar steps can be applied for topology optimization \cite{2021:Rasmus:Inverse_design_in_photonics_by_topology_optimization_tutorial}. The optimization target must be reformulated slightly since in shape optimization the target is to optimize over the possible geometrical shapes instead of the parameters. A shape $\Omega$ is a connected region within the computational domain $\mathcal{D} \subset \mathbb{R}^n$ with fixed optical material properties. The PDE can then be solved on $\mathcal{D}$ which yields solution $u_\Omega \in \mathbb{R}^k$. A loss functional $J$ is then applied to evaluate the solution. Shape optimization derives how to deform the boundary of the shape $\Omega$ to improve the loss $J(u_\Omega)= \int_{\mathcal{D}}\ds\, u_\Omega$. By taking the variation of the loss functional to first order, we obtain \cite{2019:Lebbe:Contribution_in_topological_optimization_and_application_to_nanophotonics, 2011:Jensen:Topology_optimization_for_nano-photonics, 2021:Christiansen:Inverse_design_in_photonics_by_topology_optimization:_tutorial}
\begin{align}
    \delta J_\Omega(\delta\Omega) = \int_{\p \Omega}\ds\, \delta \Omega \, \hat{n}\, \left(c_1 - c_2 \right)u_\Omega v_\Omega=  \int_{\p \Omega}\ds\, \delta \Omega \, \hat{n}\, V_\Omega(s).
\end{align}
$\delta\Omega$ denotes the variation of the shape which is equivalent to a test function in functional analysis. At iteration i, the variation deforms the shape $\Omega_{i+1} = (\mathds{1} + \delta \Omega)(\Omega_i)$. $\hat{n}$ denotes the normal vector on the boundary, and $V_\Omega(s)$ denotes the s.c. sensitivity field (also known as gradient field \cite{2013:Miller:Adjoint_shape_optimization_applied_to_electromagnetic_design} or velocity field \cite{2019:Wang:Velocity_field_level-set_method_for_topological_shape_optimization_using_freely_distributed_design_variables}). Since our goal is to minimize the loss functional, we see that this can be achieved by setting the geometry deformation to $\delta\Omega = - \hat{n}\, V_\Omega(s)$. Then, the loss functional is guaranteed to decrease to first order in every iteration. The sensitivity field acts on the shape as a vector field that drags the boundary along the direction of the vector field \cite{1991:Delfour:Velocit_Method_and_Lagrangian_Formulation_for_the_Computation_of_the_Shape_Hessian, 2019:Wang:Velocity_field_level-set_method_for_topological_shape_optimization_using_freely_distributed_design_variables}.

The forward and adjoint solutions $u_\Omega$ and $v_\Omega$ are directly inserted in the shape gradient. $c_1$ and $c_2$ are parameters of the computational domain inside and outside of the shape $\Omega$. For photonic optimization, the parameters are given by the relative electric permittivity $\varepsilon_i$ for the parallel component of the electric fields in $u$ and $v$ and $1/\varepsilon_i$ for the normal component \cite{2013:Miller:Adjoint_shape_optimization_applied_to_electromagnetic_design}. 

\begin{figure}[ht]
  \centering
  \includegraphics[angle=0, trim = 0cm 0cm 0cm 0cm, clip, height=10cm]{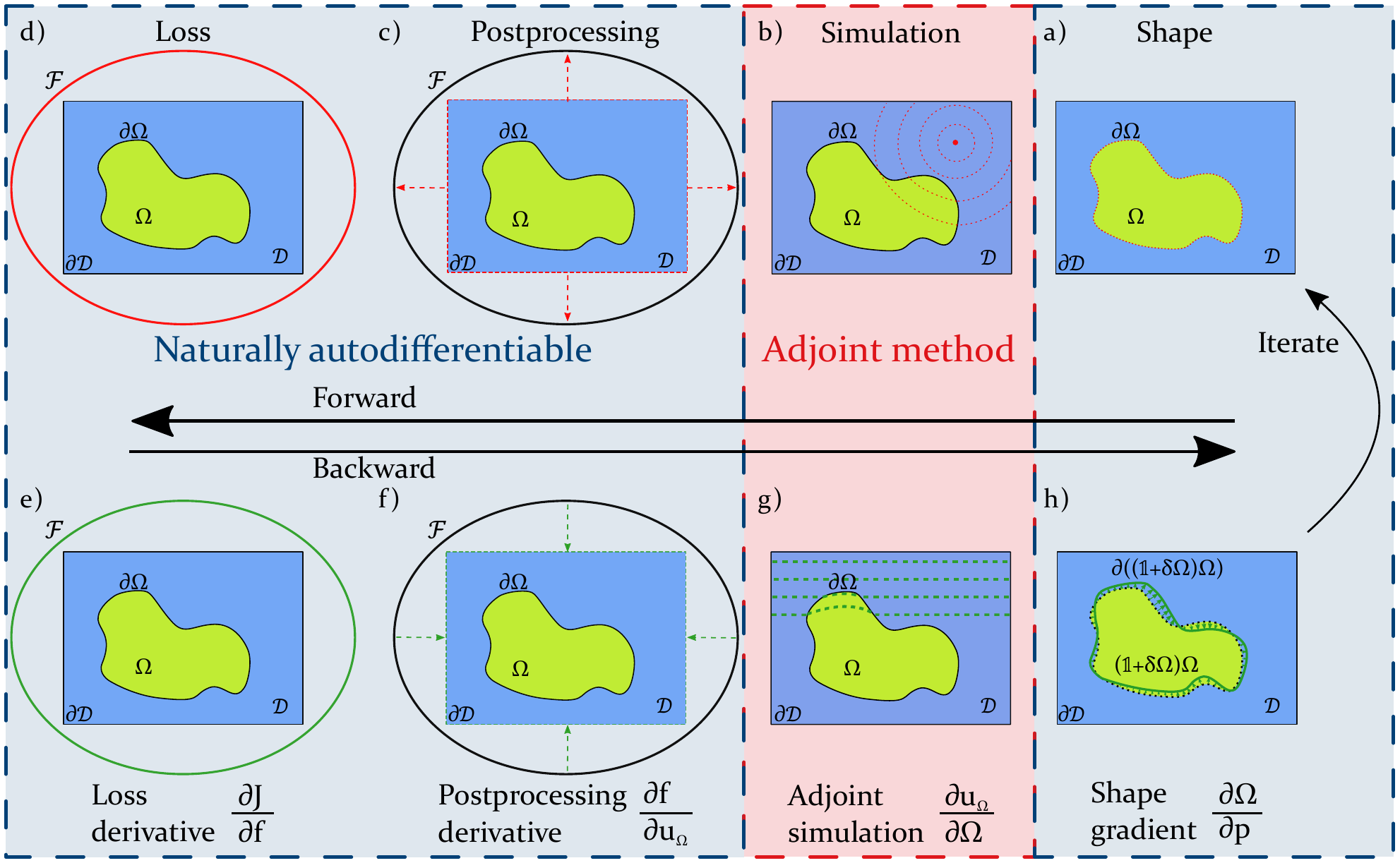}
  \caption{Overview of a typical geometry optimization problem. \textbf{a)} The optimization starts with a parametrized initial shape $\Omega(p)$ within a computational domain $\dom$. \textbf{b)} Then, the physical problem is solved using a numerical solver with the source distribution appropriate for the problem at hand. \textbf{c)} Next, the physical solution is evaluated in the domain or on its boundary and transformed during postprocessing, for example by projecting a recorded near field to the far field or selecting different waveguide modes. \textbf{d)} Finally, the result from postprocessing is evaluated and integrated by a loss functional $J$. \textbf{e-f)} Differentiating the loss functional up to the solution from the numerical solver is easy due to automatic differentiation (AD). \textbf{g)} However, it becomes more difficult to obtain gradients for the simulation parameters from the simulation solution since many numerical simulations do not provide AD or it is not efficient to use. Taking the derivative with respect to the shape parameters is therefore difficult. By combining AD with the adjoint method, backpropagation computes the gradients up to the solution from the solver at which point the adjoint method is employed to continue with the gradient computation and passes the gradients further to the shape parameters. Computing the gradient of a given shape is then reduced to backpropagating through the computational graph of the AD framework, which is equivalent to backpropagation through the simulation itself by means of the adjoint method. \textbf{h)} Finally, the gradient of the shape is evaluated by AD, and the shape parameters are updated by an optimization algorithm. Then the next iteration begins with the updated shape.
 \label{fig:sim_flow}}
\end{figure}

In many interesting scenarios, the functional $J(u_\Omega)$ has a complicated dependence on the solution from the post-processing and the shape has parameter dependencies. Generally, this postprocessing dependence is described by a function $f: \mathbb{R}^k \times \mathcal{D} \mapsto \mathcal{F}$, where $\mathcal{F}$ is a  vector space, usually chosen to be $\mathbb{R}^m$. The postprocessing, which can be arbitrarily complicated, acts on the solution $u_\Omega$ before integrating the result of the postprocessing $J(f(u_{\Omega(p)})) = \int_{\mathcal{F}}\text{dy}\, f(u_{\Omega(p)}) $ to obtain the loss $J$. An example of such a postprocessing function is the farfield transformation that is used in \autoref{sec3:µled} that projects the boundary values of the solution $u_\Omega$ from the 2D computational domain $\mathcal{D}$ on a linear farfield \cite{2011:Schneider:Understanding_the_Finite-Difference_Time-Domain_Method}. The chain rule for functionals \cite{1996:greiner:field_quantization, 2011:engles:Density_functional_theory} for chaining functionals with ordinary differentiable functions allows us to separate the postprocessing and the geometry definition from the shape gradients 
\begin{align}
    \delta J_\Omega(\delta \Omega) = \int_{\p\Omega} \ds\, \delta \Omega \, \hat{n}\, V_{\Omega(p)}(s) \frac{\p {\Omega(p)}}{\p p}.
\end{align}
The functional derivative can then be written as 
\begin{align}\label{eq:functional_derivative}
    \frac{\delta J_\Omega}{\delta p} = \hat{n}\left(c_1 - c_2 \right) u_\Omega \underbrace{\left(\frac{\p J}{\p f} \frac{\p f}{\p u_{\Omega(p)}}A^{-1}\right)}_{v_\Omega} \frac{\p \Omega(p)}{\p p}.
\end{align}
Here, we see again how to employ automatic differentiation on the adjoint source computation for the adjoint solution $v_\Omega$ and then continue with the backpropagation of the shape. For illustrative purposes, the process is also depicted in \autoref{fig:sim_flow}.

\subsection{Software considerations}
On a practical level, integrating the adjoint method into an AD framework and thus making AD compatible with conventional solvers boils down to implementing a \textit{differentiable simulation} function that receives a numerical representation of the geometry boundary $\p\Omega$. The differentiable simulation is shown in pseudocode in \autoref{appendix:differentiable_simulation_code}. Boundary support points that give rise to the shape $\Omega$ are an ideal representation since they easily integrate with numerical AD frameworks. Here, we focus only on 2D shapes but this can be extended to 3D by modifying the equations appropriately. The differentiable simulation method starts by initializing the simulation by evaluating the geometry support points. Then, the solution of the simulation is computed and returned to the AD framework. The AD framework then undertakes the postprocessing of the solution, along with the evaluation of the loss functional. Taking the derivative of the loss functional up to the adjoint simulation is handled by the AD framework using backpropagation. Crucially, the AD framework returns the \textit{adjoint source} $\frac{\p J}{\p u}$, see \autoref{fig:sim_flow}e-f), which is usually derived manually in former applications \cite{2013:Miller:Adjoint_shape_optimization_applied_to_electromagnetic_design, 2013:Lalau:Adjoint_shape_optimization_applied_to_electromagnetic_design, 2015:Hansen:Accurate_adjoint_design_sensitivities_for_nano_metal_optics, 2019:Lebbe:Contribution_in_topological_optimization_and_application_to_nanophotonics}.

The adjoint source is then passed to the \textit{backward} method of the \textit{differentiable simulation} function as the \textit{gradient input}, shown in \autoref{fig:sim_flow}g). It initializes the adjoint simulation and computes the adjoint solution $v$, which is particularly easy for Maxwell's equations with linear materials due to their time-reversal properties \cite{2015:Hansen:Accurate_adjoint_design_sensitivities_for_nano_metal_optics}. For the Maxwell equations, the adjoint system is given by $A^\dagger v_\Omega = T A T \,v_\Omega = \frac{\p J}{\p u}$ with $T$, the time reversal operator. Then, computing the adjoint solution can be done by solving $A T v_\Omega = T \frac{\p J}{\p u}$.

Taking the derivative of a function often involves retaining the results from the forward function evaluation. In the case of the adjoint method, most importantly the solution $u_\Omega$ but it is useful to also save the boundary reference and other additional parameters for the solver. Since AD frameworks are often required to store the solution of the forward pass, the framework provides the functionality to store intermediate results which are necessary for the gradient calculation during backpropagation. Together with the forward solution, the backward method computes the sensitivity field for the given geometry with the shape calculus shown in \autoref{eq:functional_derivative} \cite{2011:Delfour:Shapes_and_Geometries, 2019:Lebbe:Contribution_in_topological_optimization_and_application_to_nanophotonics,
2015:Hansen:Accurate_adjoint_design_sensitivities_for_nano_metal_optics}.

The sensitivity field will act as a gradient by deforming the geometry since $\delta\Omega$ will decrease the loss functional to first order. The movement of the boundary is projected on the normal since the tangential movement of the boundary has no influence on the loss functional which is shown in \autoref{eq:functional_derivative}. We need to take into account that the geometry is represented by boundary support points where movement of the support points drags the connecting edge along. The edge displacement must be carefully taken into account which is detailed in \autoref{appendix:support_points}. The support point sensitivities are interpreted as the support point gradients and returned by the backward method. The AD framework continues with the backpropagation to the geometry parameters $p$. In this way, it is particularly straightforward to create parameterized geometries which, for example, serve to introduce geometry constraints that ensure favorable properties such as manufacturability. To illustrate the core functionality of the \textit{differentiable simulation} function we present it in pseudo-code in \autoref{appendix:differentiable_simulation_code}.

Finally, the shape can be updated via an optimizer depicted in \autoref{fig:sim_flow}h). The optimizer uses the computed gradients in order to update the shape parameters with a stepsize estimated by the selected algorithm. Many different approaches exist and exhibit advantages and disadvantages. The simplest optimizer is gradient descent where the parameters are updated based on a stepsize $\eta$ selected at the start following the rule $p_{i+1} = p_i \pm \eta\nabla p$. However, many more refined techniques exist such as quasi-newton methods \cite{1987:Fletcher:Practical_Methods_of_Optimization} that estimate the hessian iteratively with gradient information or moment estimation methods such as ADAM \cite{2017:kingma:Adam:_A_Method_for_Stochastic_Optimization} which approximate first and/or second-order moments of the stepsize.


\section{Application examples}
\label{sec:4}
To showcase the integration of a conventional simulation into an AD framework, we apply it to two different problems of current interest. In the first example, we enhance the spontaneous emission rate for an optical nanocavity while in the second example, we increase the farfield intensity distribution within a given solid angle by optimizing the outcoupling structure of a 2D µLED.

\begin{figure}[ht]
\centering
  \includegraphics[angle=0, trim = 0cm 0cm 0cm 0cm, clip, width=0.8\textwidth]{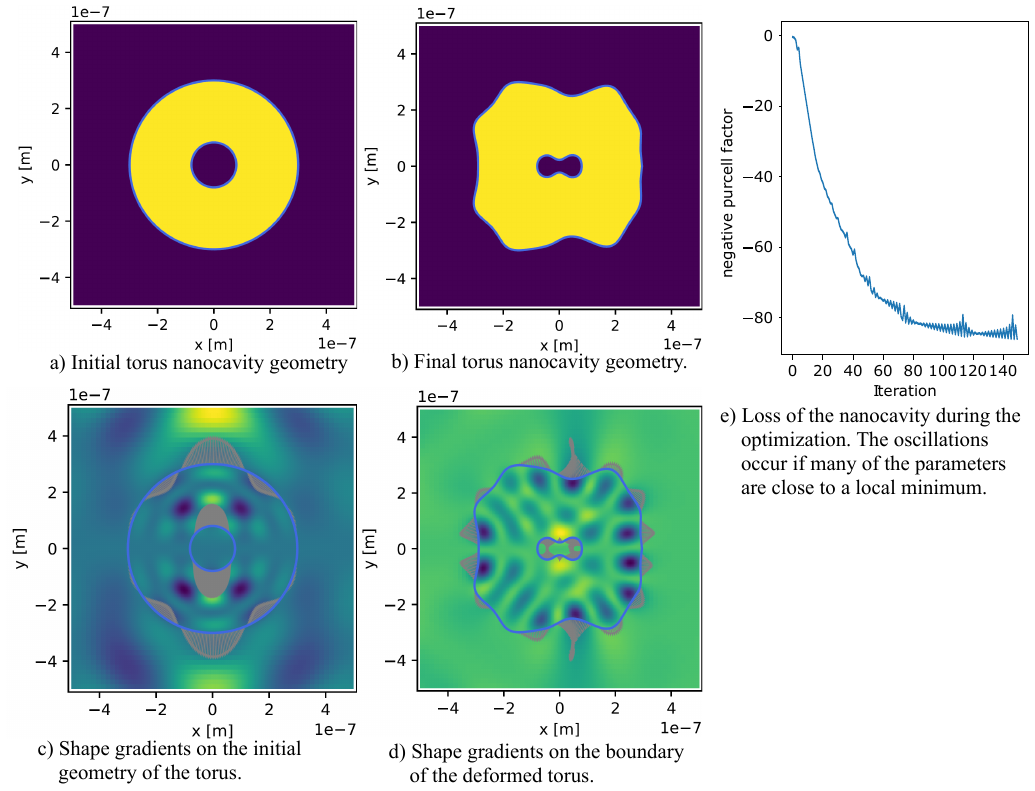}
\caption{Overview of the gradient descent optimization results for the enhancement of the Purcell factor in a torus nanocavity. \textbf{a)} depicts the initial torus geometry. The dipole emitter sits in the center of the torus at $\Vec{r} = (0,0)^T$. The refractive index of the torus has a value of $n=3.5+0i$. The blue contour represents the deformable boundary which is subject to optimization. The optimized torus geometry after 150 iterations is shown in \textbf{b)}. Geometry deformations during optimization led to a loss reduction, as demonstrated in \textbf{e)}. The loss exhibits a steady decline throughout the optimization process with slight oscillations which emerge if large parts of the boundary are close to a local minimum. For this optimization, we chose the negative Purcell factor for the loss with gradient descent. After the optimization, the Purcell factor was increased by about $\times 220$. The deformations are computed via the adjoint method, see \autoref{sec:adjoint_method_AD}. The sensitivity field is shown in \textbf{d)} together with the shape gradients that are obtained by evaluating the sensitivity field on the boundary of the deformable geometry. 
\label{fig:purcell_optimization}}
\end{figure}

\subsection{Purcell effect optimization}\label{sec3:purcell}
In the example shown in \autoref{fig:purcell_optimization}, we aim to increase the spontaneous emission rate for an optical nanocavity. Increasing the spontaneous emission rate is a research problem both for inverse design \cite{2011:Lu:Inverse_design_of_a_three-dimensional_nanophotonic_resonator, 2010:Lu:Inverse_design_of_nanophotonic_structures_using_complementary_convex_optimization} and classical approaches \cite{2019:Mignuzzi:Nanoscale_Design_of_the_Local_Density_of_Optical_States}. More precisely, our goal is to increase the Purcell factor, which is proportional to the time-averaged poynting vector on the boundary of the domain $\max \frac{P}{P_0} =\max \int_{\p\dom} \bold{\hat{n}}\Re\bigl[\bold{E}_\Omega \times \bold{H}_\Omega^\dagger\bigr]/2 \, \ds$ where $P_0$ is the dipole emission in free space, which is used for normalization while $\hat{n}$ denotes the boundary normals and $\bold{E}_\Omega$ and $\bold{H}_\Omega$ denote electric and magnetic fields on the boundary $\Omega$. We employ the FDTD solver from Lumerical \cite{lumerical} to obtain the electric and magnetic field solutions for the selected wavelength of 600nm. For simplicity and to reduce the simulation time, we consider a 2D problem that is infinitely extended in the z-direction. In the center of the simulation domain, we place a magnetic dipole emitter with the magnetic current oriented in the x-direction and surround it with a torus structure with a real refractive index of $n=3.5$. Both the inner and the outer torus boundary, shown in \autoref{fig:purcell_optimization}-, are subject to optimization but are represented by a tensor of an AD framework. In the presented case, we employ PyTorch as AD framework. For the update scheme, we employ simple gradient descent but chose a physically inspired step size on a length scale for which we expected a change of the Purcell factor. We also decrease the step size proportionally to the Purcell factor. The optimization shows converging behavior after 150 iterations, in which the Purcell factor increased to $\frac{P}{P_0} \approx 83$ starting from $\frac{P}{P_0} \approx 0.4$, which is in the range of expected improvement. Since we use a simple gradient update scheme, the presented solution is highly dependent on the initial geometry and is potentially far away from a global maximum for the Purcell factor. The optimization also results in a dumbbell shape around the dipole which has been observed in other works, too \cite{2019:Mignuzzi:Nanoscale_Design_of_the_Local_Density_of_Optical_States, 2017:Wang:Optimization_of_photonic_crystal_cavities, 2011:Lu:Inverse_design_of_a_three-dimensional_nanophotonic_resonator, 2021:Yesilyurt:Efficient_Topology-Optimized_Couplers_for_On-Chip_Single-Photon_Sources, 2010:Lu:Inverse_design_of_nanophotonic_structures_using_complementary_convex_optimization}.

\subsection{Shape optimization for spatially distributed dipole emission of a µLED}\label{sec3:µled}
\begin{figure}[H]
\centering
  \includegraphics[angle=0, trim = 0cm 0cm 0cm 0cm, clip, width=0.99\textwidth]{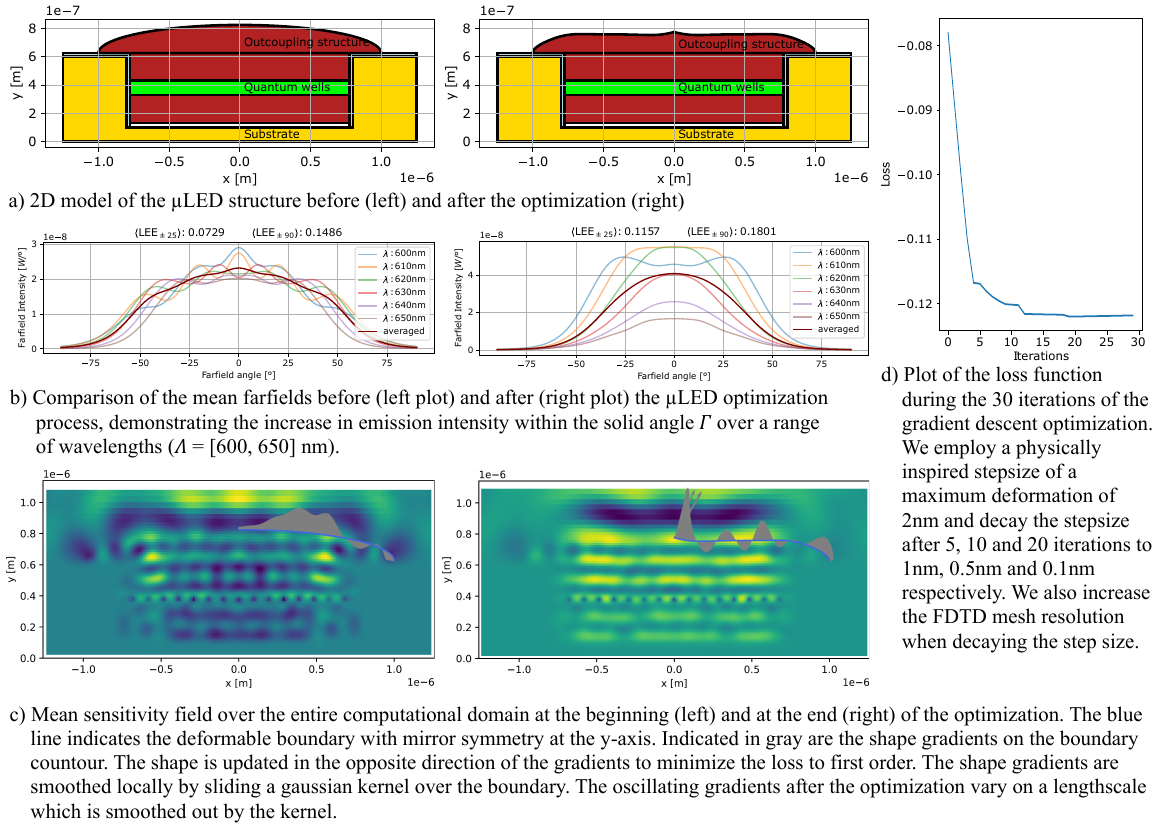}
    \caption{Illustration of the µLED optimization process: \textbf{a)} presents the initial model, where dipole emitters are distributed within the quantum well region (green). The averaged farfield emission of the initial µLED and LEE into solid angle $\Gamma=25^\circ$ is shown on the left plot of \textbf{b)}. For reference, \textbf{b)} also displays the farfield for a range of individual wavelengths and the total LEE. Using the adjoint method with shape calculus \autoref{sec:adjoint_method_AD} we obtain the averaged sensitivity field over the entire domain, which is shown in \textbf{c)}. By following the direction of the steepest descent and updating the boundary geometry according to the shape gradients for 30 iterations, we deform the boundary and decrease the loss over the optimization duration, see \textbf{d)}. To avoid obtaining a shape with very small features, we smooth the gradients locally over the deformable boundary. At the end of the optimization, the outcoupling structure of the µLED is shown in \textbf{a)} on the right. The corresponding farfield for the optimized µLED shown on the right in \textbf{b)} experiences a reshaping with an improvement of the overall LEE directed into $\Gamma$ of $\Delta\text{LEE}_{\pm 25} = 0.0428$ and a total $\text{LEE}$ improvement of $\Delta\text{LEE}_{\pm 90} = 0.0315$. }
    \label{fig:sandbox_µLED}
\end{figure}

In the example shown in \autoref{fig:sandbox_µLED}, our primary objective is to enhance the farfield emission of a µLED (nanoscale light-emitting diode) within a specific solid angle, denoted as $\Gamma$, also known as LEE (light extraction efficiency). The field of µLED development is rapidly advancing, attracting substantial research interest and holding considerable industrial relevance \cite{2023:Li:Significant_Quantum_Efficiency_Enhancement_of_InGaN_Red_Micro-Light-Emitting_Diodes_with_a_Peak_External_Quantum_Efficiency_of_up_to_6, 2021:Li:High-temperature_electroluminescence-properties_of_InGaN_red-40x40µm²-micro-light-emitting_diodes_with_a_peak_external_quantum_efficiency_of_3.2, 2019:Taki:Visible_LEDs:More_than_efficient_light, 2022:Chung:Computational_upper-limit_of_directional_light_emission_in_nano-LED_via_inverse_design}. The targeted optimization can be mathematically represented by the function $\max_\Omega \text{LEE}_\Gamma = \max_\Omega  P_\Gamma/P_0=\max_\Omega 
\frac{1}{P_0}\int_{\Gamma} \int_\Lambda \Re\bigl[E_{\Omega,\,\text{Farfield}} \times H^\dagger_{\Omega,\,\text{Farfield}}\bigr]/2\, \d\Gamma \d\lambda$. LEDs exhibit incoherent dipole emissions that originate from all over the quantum well region. To approximate this emission behavior, we distribute dipole emitters over the quantum well and compute the average emitted intensity within the solid angle $\Gamma$ and normalize with respect to the injected power $P_0$ into the simulation. By solving the emission problem for each individual dipole, we can account for the incoherent nature of the µLED's emission. To simplify the problem and reduce computational time, we adopt a two-dimensional FDTD model. The wavelength range of interest for this example is from 600 to 650 nanometers ($\Lambda = [600, 650]$ nm). The initial µLED model is depicted in \autoref{fig:sandbox_µLED}a). The µLED features a gold substrate passivated with a thin layer of silicon dioxide and a semiconductor material composed of gallium nitride, which is connected to the gold substrate via a thin layer of indium tin oxide. The quantum well region is made up of indium gallium phosphide. To enhance the light extraction efficiency within the solid angle $\Gamma$, we deform the upper boundary of the µLED's top side, referred to as the outcoupling structure, which consists of gallium nitride. The emission then radiates into the air.

To enable the use of automatic differentiation, we represent the outcoupling structure geometry with a PyTorch tensor. The differentiable simulation receives a reference to this boundary representation and creates an STL file from the tensor that is imported to the solver after each iteration. Furthermore, the projection of the fields on the boundary of the µLED are projected into the farfield by a PyTorch implementation of the equivalence principle in order to use AD for the postprocessing
\cite{2011:Schneider:Understanding_the_Finite-Difference_Time-Domain_Method}. Employing a standard gradient descent optimization technique, we run the optimization for 30 iterations. During the optimization, the outcoupling structure is continuously deformed following the mean sensitivity field computed by the adjoint method. The shape gradients and mean sensitivity field for the initial and final structure are shown in \autoref{fig:sandbox_µLED} a) in the left and right figure respectively. The boundary approximation of the outcoupling structure is much finer than the mesh size of the solver, thus we interpolate the field solution to obtain smooth field values on the boundary. For optimization stability, we decay the stepsize after 5, 10, and 20 iterations from 2nm to 1nm, 0.5nm, and 0.1nm. By increasing the mesh resolution, after 10 and 20 iterations, we make sure to resolve the geometry changes sufficiently when the stepsize becomes small. To avoid creating small features on the outcoupling structure boundary, we smooth the shape gradients locally by sliding a Gaussian kernel over the boundary. Therefore, the standard deviation of the kernel controls the size of the features. This explains the oscillating gradients on the boundary after the optimization ( see \autoref{fig:sandbox_µLED} c) plot on the right), which are smoothed out after applying the Gaussian kernel. The final optimized µLED structure is presented in \autoref{fig:sandbox_µLED} a), while the loss throughout the optimization process is depicted in \autoref{fig:sandbox_µLED} d). As the main objective is proportional to the farfield intensity within the solid angle $\Gamma$, we also provide the mean farfield at both the beginning and end of the optimization in \autoref{fig:sandbox_µLED} b). Additionally, we show the farfield intensity for a range of wavelengths. After the optimization, the averaged farfield is focused in the target solid angle $\Gamma$, and the LEE$_\Gamma$ is increased by 0.0428, an improvement of $63.01\%$ compared to the initial farfield.

\section{Conclusion}
\label{sec:5_conclusion}

In this work, we have presented an general approach to include existing numerical solvers into an automatic differentiation framework to simplify and make the optimization of photonic structures faster and more convenient, particularly with existing models and solvers. To this end, the adjoint method is the key to implementing the forward and backward methods for automatic differentiation which allows us to make conventional solvers end-to-end auto-differentiable. Due to our focus on continuous geometries, we presented the optimization in the context of shape optimization and computed shape gradients on geometry boundaries which enable gradient-based optimization algorithms to improve the optical characteristics of the geometry with respect to the loss function. Generally, the approach is also suitable for topology optimization.

We demonstrate the approach with two different physical problems, optimizing an optical nanocavity to increase the spontaneous emission rate and optimizing the outcoupling structure of a µLED to increase the light extraction efficiency into a solid angle in the farfield. For both cases, we show a significant reduction of the loss while employing Lumerical FDTD to solve the Maxwell equations.

Two key advantages of employing automatic differentiation (AD) for numerical optimization are parallelized multiphysics optimization and compatibility with machine learning. By integrating various physics solvers into AD, conducting joint optimization (e.g., thermal and optical optimization) becomes considerably more straightforward. Integrating the adjoint method into AD frameworks could also facilitate the development of novel AI applications for designing and optimizing optical devices. 


\subsection*{Acknowledgments}
We thank the supporters of this work, particularly Heribert Wankerl (ams-OSRAM Group) and Maike Stern (OTH Regensburg) for constructive feedback of this manuscript. Furthermore, we thank Daniel Grünbaum (ams-OSRAM Group) and Philipp Schwarz (ams-OSRAM Group \& University Regensburg) for fruitful and helpful discussions. Finally, we thank Harald Laux (ams-OSRAM Group) for his organizational support.

\subsection*{Publication Funding Acknowledgments}
This work was funded by the ams-OSRAM Group.







\bibliographystyle{unsrt}  
\bibliography{0_references}  

\section{Appendix}
\label{sec:appendix}
\subsection{Support Points behavior}
\label{appendix:support_points}
\begin{figure}[ht]
      \centering
      \includegraphics[angle=0, trim = 0cm 0cm 0cm 0cm, clip, height=2.2cm]{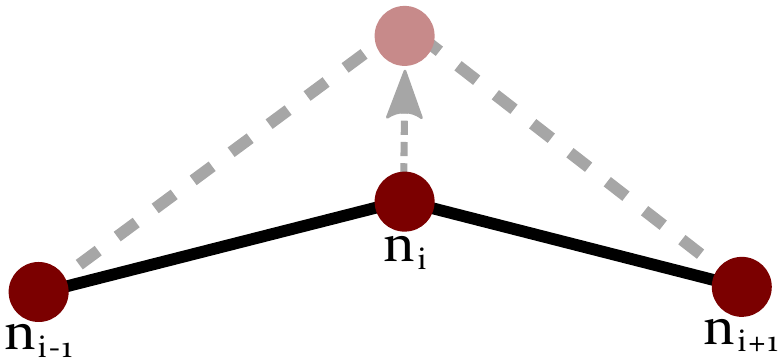}
      \caption{Edge movement for a displacement of an individual support point. The edge moves much further close to the support point $n_i$ than close to the adjacent support points. For a small displacement of support point $n_i$, the variation of the support point is given by \autoref{eq:support_point_movement}. \label{fig:support_point_movement}}
\end{figure}

The representation of the shape with support points is quite important for the gradient computation. The shape of the discretized geometries are deformed by dragging the support points following the gradients. First of all, the gradient in shape optimization is only meaningful in direction of the normal of the boundary. The gradient direction is therefore a weighted combination of the normal vectors of the two adjacent edges to the support points. Furthermore, depending on the density of the support points, one can not simply evaluate the sensitivity field on the support points position $x_{n_i}$ directly since the sensitivity field can vary between support points. Moving an individual support point leads to moving the edge between the previous support point $x_{n_{i-1}}$ and the next support point $x_{n_{i+1}}$ see \autoref{fig:support_point_movement}. The influence of a small displacement of an individual support point $n_i$ on the variation of the loss functional can be computed by \cite{2013:Miller:Adjoint_shape_optimization_applied_to_electromagnetic_design}
\begin{align}\label{eq:support_point_movement}
    \delta n_i = \int_{x_{n_{i-1}}}^{x_{n_i}} \dx \frac{x - x_{n_{i-1}}}{x_{n_i}} \hat{n} V_\Omega(x) - \int_{x_{n_i}}^{x_{n_{i+1}}} \dx \frac{x - x_{n_{i}}}{x_{n_{i+1}}} \hat{n} V_\Omega(x).
\end{align}

\subsection{Pseudo-code for the differentiable simulation function}\label{appendix:differentiable_simulation_code}
The forward and backward methods seamlessly integrate into the computational graph of the chosen autograd framework where using forward and backward is often the naming convention. The inputs to the forward are necessary to specify the boundary conditions for the numerical solution. For Lumerical, this would mean activating the forward sources for the forward calculation and activating the adjoint sources for the backward calculation.   After the simulation, the numerical solution from the simulation is collected and passed to the next function in the computational graph. After calling backward on the loss value and backpropagating up to the \textit{differentiable simulation}, the adjoint simulation is initialized with the adjoint sources and the given shape. Note that the time reversal operation which is performed on the gradient input and the adjoint solution. The forward and adjoint solutions are then used to compute the sensitivity field \cite{2019:Lebbe:Contribution_in_topological_optimization_and_application_to_nanophotonics}. c1 and c2 are constants, which depend on the permeability of the materials inside and outside of the boundary. Finally, the sensitivity field is evaluated on the boundary and projected on the boundary normal. These values then represent the gradients on the shape which are returned from to the computational graph.

\begin{algorithm}[ht]
        \begin{algorithmic}[1]
            \Class{DifferentiableSimulation(autograd)}
            \\
                \Function{forward}{self, simulation, forward\_source, c1, c2, shape}
                    \State self.store $\gets$ simulation
                    \State self.store $\gets$ c1, c2
                    \State self.store $\gets$ shape
                    \State simulation $\gets$ forward\_Source
                    \State simulation $\gets$ shape
                    \State u $\gets$ simulation.run()
                    \State self.store $\gets$ u
                    \State \Return u
                \EndFunction
            \\
                \Function{backward}{self, grad\_input}
                    \State simulation $\gets$ self.store
                    \State shape $\gets$ self.store
                    \State adjoint\_source $\gets$ time\_reverse(grad\_input)
                    \State simulation $\gets$ adjoint\_source
                    \State simulation $\gets$ shape
                    \State v $\gets$ simulation.run()
                    \State u $\gets$ self.store
                    \State c1, c2 $\gets$ self.store
                    \State sensitivity\_field $\leftarrow$ u, v, c1, c2
                    \State V\_Omega $\gets$ sensitivity\_field(shape)
                    \State normals $\gets$ shape
                    \State gradients $\gets$ normals $\cdot$ V\_Omega
                    \State \Return gradients
                \EndFunction
            \EndClass
        \end{algorithmic}

    \caption{Pseudocode for a \textit{differentiable simulation} autograd function for automatic differentiation to derive shape gradients via the adjoint method. }
    \label{algo:differentiable_simulation}
\end{algorithm}

\end{document}